\documentclass{iaus}
\usepackage{graphicx,natbib}

\newcommand{\sauron}{\texttt{SAURON}}
\newcommand{\oasis}{\texttt{OASIS}}
\newcommand{\mnras}{MNRAS}
\newcommand{\apj}{ApJ}
\newcommand{\apjl}{ApJ}

\newcommand{\aap}{A\&A}
\newcommand{\nat}{Nature}

\title[Black holes with OASIS]
{Supermassive black holes from OASIS and SAURON integral-field kinematics}

\author[Cappellari et al.]{
	Michele Cappellari,$^1$
	R.\ Bacon,$^2$
	Roger L.\ Davies,$^1$
	P.\ T.\ de Zeeuw,$^3$
	Eric Emsellem,$^2$
	Jes\'us Falc\'on-Barroso,$^4$
	Davor Krajnovi{\'c},$^1$
	Harald Kuntschner,$^5$
	Richard M.\ McDermid,$^3$
	Reynier F.\ Peletier,$^6$
	Marc Sarzi,$^7$
	Remco C. E. van den Bosch,$^3$
	Glenn van de Ven$^8$}

\affiliation{
    $^1$Sub-Department of Astrophysics, University of Oxford, England \\ [\affilskip]
    $^2$Univ.\ de Lyon~1, CRAL Observ. de Lyon; CNRS, Ecole Normale Sup\'erieure de Lyon, France \\ [\affilskip]
    $^3$Sterrewacht Leiden, Leiden University, Leiden, The Netherlands \\ [\affilskip]
    $^4$European Space and Technology Centre, Noordwijk, The Netherlands \\ [\affilskip]
    $^5$Space Telescope European Coordinating Facility, ESO, Garching, Germany\\ [\affilskip]
    $^6$Kapteyn Astronomical Institute, Groningen, The Netherlands \\ [\affilskip]
    $^7$Centre for Astrophysics Research, University of Hertfordshire, England \\ [\affilskip]
    $^8$Institute for Advanced Study, Princeton, USA}

\pubyear{2007}
\volume{245}
\jname{Formation and Evolution of Galaxy Bulges}
\editors{M. Bureau, E. Athanassoula, and B. Barbuy, eds.}
\begin{document}

\maketitle

\begin{abstract}
Supermassive black holes are a key element in our understanding of how galaxies form. Most of the progress in this very active field of research is based on just $\sim30$ determinations of black hole mass, accumulated over the past decade. We illustrate how integral-field spectroscopy, and in particular our \oasis\ modeling effort, can help improve the current situation.
\end{abstract}

\firstsection
\section{Supermassive black holes and galaxy evolution}

Fifteen years ago the existence of supermassive black holes (BHs) in galaxy nuclei was considered an interesting possibility which had to be demonstrated. Nowadays BHs are regarded as the crucial ingredient for our understanding of how galaxies form. Key to this paradigm shift was the launch in 1990 of the Hubble Space Telescope (HST). It all started with the realisation that the mass of the BH is correlated to other global characteristics of the host galaxy as a whole. Initially a correlation  $M_{\rm BH}-L$ was found between the mass of the BH and the luminosity of the host-galaxy stellar spheroid \citep{kor95,mag98}. In 1997 the installation of the STIS long-slit spectrograph on HST allowed the spatially-resolved kinematical observations to probe inside the radius of the subarcsecond BH sphere of influence $R_{\rm BH}\equiv G M_{\rm BH}/\sigma^2$ in nearby galaxies ($\sigma$ being the velocity dispersion of the galaxy stars). The increased accuracy in the $M_{\rm BH}$ determinations contributed to the discovery of the much tighter $M_{\rm BH}-\sigma$ correlation \citep{geb00,fer00}.

Similar correlations were found between the $M_{\rm BH}$ and respectively the galaxy concentration \citep{gra01}, the dark-halo mass \citep{2002ApJ...578...90F,2005ApJ...631..785P}, the galaxy mass \citep{2003ApJ...589L..21M,2004ApJ...604L..89H} and the stars gravitational binding energy \citep{2007ApJ...665..120A}. The existence of these correlations is broadly consistent with a scenario in which the BH regulates the galaxy formation, during the hierarchical galaxy merging, by shutting off the conversion of gas into stars via a feedback mechanism due to its powerful jet \citep{1998A&A...331L...1S,2005Natur.433..604D}.

\section{Observational evidences and current limitations}

\begin{figure}
  \centering
  \includegraphics[width=0.55\textwidth]{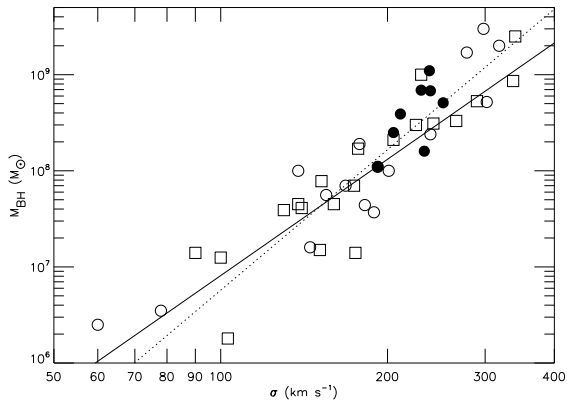}
  \caption{Updated $M_{\rm BH}-\sigma$ relation. Open squares are values from the literature; open circles are literature values, plotted against the luminosity-weighted $\sigma$ within 1$R_e$ determined from \sauron\ data \citep[see][]{cap06}; filled circles are our new $M_{\rm BH}$ \oasis\ determinations, against the \sauron\ $\sigma$. For reference, the solid line is the relation of \citet{2002ApJ...574..740T}, while the dotted one is from \citet{2005SSRv..116..523F}.}
  \label{fig:bh-sigma}
\end{figure}

Our understanding of the role of BHs in galaxy evolution is however far from complete. In fact the current observables do not uniquely constrain the models, which depends on a number of assumptions. This is due in part to the relatively small number of secure BH measurements: in a decade of high-resolution observations and models only $\sim30$ values have been obtained \citep[see][for a recent review]{2005SSRv..116..523F}. It is remarkable that so much progress was based on an extrapolation of so few BH measurements, and for a biased galaxy sample, to the entire galaxy population!

An additional complication comes from the fact that the above correlations provide a rather indirect test for the models. A complementary and more direct approach to test the BH formation paradigm consists of looking in nearby galaxy centres for the signatures of the joint formation of the BH and the galaxy spheroid. Simulations of galaxy mergers show in fact that, when two galaxies with a BH merge, the distribution of the stellar orbits in the resulting remnant, after the two BHs coalesce, is significantly different from the one of the progenitor galaxies. This is due to the ejection of stars passing, along radial orbits, close to the resulting BH binary. Two observables signatures are expected: (i) the density profile should flatten inside the core radius $R_C$, which is much larger than $R_{\rm BH}$ and (ii) the orbital distribution should be biased towards tangential orbits inside $R_C$ \citep{1997NewA....2..533Q,2001ApJ...563...34M}.

Evidence for the formation of the density core, and its expected relation to the BH mass, was found from photometric observations \citep{fab97,2002MNRAS.331L..51M}. The detection of the orbital signature is more complicated as it requires integral-field spectroscopic observations at high resolution \citep[e.g.][]{cmd05}. The orbital distribution, in a non-spherical stationary system, is in fact a function of the three isolating integrals of motion and requires at least a three-dimensional observable quantity to be constrained. Nonetheless a first attempt at deriving the nuclear stellar anisotropy, for a sample of 12 galaxies, was done using long-slit STIS spectroscopy by \citet{2003ApJ...583...92G}. They found a tendency for the orbits of the most massive objects to be tangentially biased. However this appears to be true only well inside $R_{\rm BH}$ and not up to $R_C$ as predicted by the simulations. Similar results were recently found for another carefully studied galaxy by \citet{2006MNRAS.367....2H} and \citet{geb07}.

\section{The role of \oasis\ integral-field spectroscopy}

\begin{figure}
  \includegraphics[width=\textwidth]{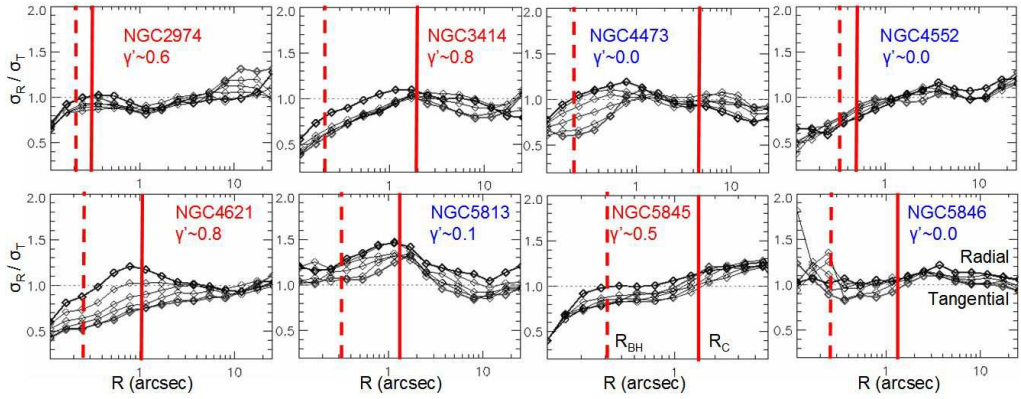}
  \caption{Anisotropy profiles from the dynamical models. In standard spherical coordinates $\sigma_R$ is the second moment of the velocity distribution along the radial direction, while $\sigma_T^2=(\sigma_\theta^2+\sigma_\phi^2)/2$. Different lines represent measurements at equally-spaced radial sectors in the galaxy meridional plane, from the equatorial plane to the symmetry axis. The vertical dashed and solid thick line indicate the position of $R_{\rm BH}$ and $R_C$ respectively. The values of $R_C$ and the logarithmic nuclear slope  $\gamma'$ in the surface brightness were taken from \citet{2007ApJ...664..226L}.}\label{fig:anisotropy}
\end{figure}

Ground-based and hig-$S/N$ integral-field observations of the stellar kinematics can overcome the limitations discussed in the previous section, and can be obtained for large galaxy samples using large telescope mirrors. By tightly constraining the orbital structure, dynamical models fitted to integral-field observations (i) can measure BH masses with accuracy comparable or better than that obtained with HST spectroscopy, even when $R_{\rm BH}$ is not well resolved \citep{sha06}; (ii) do not suffer from the degeneracy in the recovery of the orbital distribution.

For this we observed with the \oasis\ integral-field spectrograph a sample of 28 elliptical and lenticular galaxies \citep{2006MNRAS.373..906M}. The galaxies were selected from the \sauron\ sample \citep{2002MNRAS.329..513D}, for which the needed large-scale integral-field kinematics is also available up to about one half-light radius $R_e$ \citep{2004MNRAS.352..721E}. The \oasis\ observations complement the \sauron\ ones by providing an order of magnitude increase in the pixels density and a factor of two improvement in the median seeing, resulting in subarcsecond resolution.

Here we report some results of the stellar dynamical models for an initial set of eight galaxies from the \oasis\ sample. We constructed the models using our axisymmetric implementation \citep{cap06} of the orbital-superposition method \citep{1979ApJ...232..236S}, and we combined the \sauron\ and \oasis\ kinematics as in \citet{sha06}. We find that the $M_{\rm BH}$ is recovered with a median {\em formal} error of 30\% (at the 3$\sigma$ confidence level, for one degree of freedom, i.e.\ $\Delta\chi^2=9$). This accuracy is similarly to that of the previous HST determinations using long-slit spectroscopy. Moreover our integral-field kinematics allows for a more rigorous and robust determination of the luminosity-weighted second velocity moment $\sigma$ inside $R_e$, which is used in the $M_{\rm BH}-\sigma$ diagram. Our new measurements appear to follow within the scatter of the previous values (Fig.~\ref{fig:bh-sigma}).

The integral-field spectroscopy allows for a robust recovery of the nuclear orbital distribution. Our first results do {\em not} show any evidence for the predicted connection between the existence and strength of the density core, and the presence of tangentially biased orbits within $R_C$  (Fig.~\ref{fig:anisotropy}). Galaxies with a flat core do not necessarily show an increase of the tangential orbits inside  $R_C$. Tangential anisotropy is generally found only near the smaller $R_{\rm BH}$, at the limit of our spatial resolution. Our recovered anisotropy is more robust and appears less noisy than that derived from previous models based on long-slit observations, but our finding is not inconsistent with them.

A statistically more significant sample, with models based on integral-field observations, is needed to confirm the apparent discrepancy between the observations and the predictions of the binary-BH core-scouring mechanism. Additional accurate BH determinations are needed to explore the dependence,  of the various correlations involving BHs, on other galaxy global parameters. Our \oasis\ modeling effort provides a step in this direction. However higher-resolution integral-field observations, as can be obtained with the adaptive optics technique \citep[e.g.][]{2007MNRAS.379..909N}, are needed to study less massive galaxies, and to probe deeper inside $R_{\rm BH}$. At the same time new and more realistic N-body simulations, in a cosmological context which includes the hierarchical merging process, are required to be compared with the observations. This will ultimately confirm our current understanding of the role of BHs in galaxy formation or require a revision of the picture we constructed in the past decade.

\end{document}